\documentclass[preprint,12pt]{elsarticle}

\usepackage{amssymb}
\usepackage{amsmath}
\usepackage{amsthm}
\usepackage{url}
\usepackage{bbm}

\usepackage[tikz]{bclogo}
\usepackage{xcolor}
\usepackage{dsfont}
\usepackage{tkz-graph}
\usepackage{tikz}
\usepackage{tikz-3dplot}

\usepackage{comment}
\usepackage{subcaption}
\usepackage{hyperref}

\usepackage{algorithm}
\usepackage{algpseudocode}
\usepackage{tcolorbox}

\newtheorem{defi}{Definition}

\usepackage{dsfont}
\usepackage{stmaryrd}

\def\shuffle{\sqcup\mathchoice{\mkern-7mu}{\mkern-7mu}{\mkern-3.2mu}{\mkern-3.8mu}\sqcup}

\journal{Mathematics and Computers in Simulation.}

\begin{document}

\begin{frontmatter}



\title{Detecting malignant dynamics on very few blood samples using signature coefficients} 


\author{Rémi Vaucher\corref{cor1}\fnref{label1,label2}}
\author[label1]{Stéphane Chrétien}
\cortext[cor1]{Corresponding author: vaucher.maths@gmail.com}

\affiliation[label1]{organization={Laboratoire ERIC, Université Lyon 2},
            addressline={ 5 Av. Pierre Mendès France}, 
            city={Bron},
            postcode={69500},
            country={France}}
\affiliation[label2]{organization={Halias Technologies},
            addressline={27 Bd des Alpes}, 
            city={Meylan},
            postcode={38240},
            country={France}}

\begin{abstract}
Recent discoveries have suggested that the promising avenue of using  circulating tumor DNA (ctDNA) levels in blood samples provides reasonable accuracy for cancer monitoring, with extremely low burden on the patient's side. It is known that the presence of ctDNA can result from various mechanisms leading to DNA release from cells, such as apoptosis, necrosis or active secretion. One key idea in recent cancer monitoring studies is that monitoring the dynamics of ctDNA levels might be sufficient for early multi-cancer detection. This interesting idea has been turned into commercial products, e.g. in the company named \href{https://grail.com/}{GRAIL}. 

In the present work, we propose to explore the use of Signature theory for detecting aggressive cancer tumors based on the analysis of blood samples. Our approach combines tools from continuous time Markov modelling for the dynamics of ctDNA levels in the blood, with Signature theory for building efficient testing procedures. Signature theory is a topic of growing interest in the Machine Learning community \cite{chevyrev2016primer,fermanian2021learning}, which is now recognised  as a powerful feature extraction  tool for irregularly sampled signals. The method proposed in the present paper is shown to correctly address the challenging problem of overcoming the inherent data scarsity due to the extremely small number of blood samples per patient. The relevance of our approach is illustrated with extensive numerical experiments that confirm the efficiency of the proposed pipeline.
\end{abstract}



\end{frontmatter}



\section{Introduction}

Early detection of aggressive cancer tumors using the ctDNA levels in blood samples has been a topic of extensive research lately and represents a potential game changer in this area, with high expected societal impact. Previously used methods for cancer investigation include heavy use of complex imaging techniques as well as biopsy, which, despite their efficacy, come with invasiveness, high overall costs. Moreover, most methods for assessing the possible presence of a malignant cancer tumor are used after the signs of a change in the patient's health become visible to the patient, hence facing the risk of treating the disease irredeemably late. Fortunately, recent advances in molecular biology and bioinformatics have opened up new avenues for non-invasive cancer detection, based on novel types of analysis of relevant biomarkers present in blood samples. In particular, an important recent breakthrough for diagnosis and monitoring is the circulating tumor DNA (ctDNA), and recent technologies have emerged that leverage such biomarkers to achieve unprecedented performance in multi-cancer early detection. In practice, however, accurately detecting the potential signals of malignancy poses a difficult challenge for machine learning models, due to the scarsity of the observed data in real life settings. Recent analysis of the techniques based on ctDNA levels measurements revealed further need for reliable methods for  controlling the False Positive rate in particular.

The present work proposes a new approach to early cancer detection from only a few blood samples using the recent theory of Signatures \cite{chevyrev2016primer}. The Signature transform is known in particular for efficiently extracting and amplifying the distinctive features of signals. Our proposed approach relies on computing the Signature transform of the evolution of ctDNA levels. When these are modelled using a specific Birth and Death Process governing the  shedding of ctDNA biomarkers in the blood stream, a statistical test can be devised for delineating between benign tumors and malignant tumors. 

Signature Theory is one of the recent fruits of a research trend which started in Control Theory in the 60's \cite{chen1973iterated} and which reshaped Stochastic Analysis under the impulse of Terry Lyons' work \cite{friz2010multidimensional}, \cite{baudoin2014diffusion}. Signature theory was recently applied to machine learning \cite{chevyrev2016primer,fermanian2021learning} and was shown to achieve state of the art prediction performance without requiring any training in the feature extraction phase. It is widely believed that Signatures contain meaningful nonlinear information about the signals they are computed from, and that they seamlessly adapt to irregular sampling, a key advantage in our envisaged application to blood samples analysis where observations arrive at random times. Our main contribution is to show that the Signature Transform can be put to work in building an efficient pipeline for  testing the presence of malignant cancer tumor growth in a patient. We study the performance of our approach using numerical experiments that can be reproduced using the codes available at \url{https://github.com/RemiVaucher/Thesis/tree/main/Chapter4/Oncology}. 

\begin{tcolorbox}[title=Summary of Notation,colback=gray!5!white,colframe=gray!75!black]
\begin{center}
\begin{tabular}{ll}
\hline
\textbf{Symbol} & \textbf{Description} \\
\hline
$A_t$ & Number of malignant cells at time $t$ \\
$B_t$ & Number of benign cells at time $t$ (assumed constant) \\
$C_t^A$ & Amount of ctDNA in bloodstream from malignant\\ & cells at time $t$ \\
$C_t^B$ & Amount of ctDNA in bloodstream from benign cells\\ & at time $t$ \\
$C_t = C_t^A + C_t^B$ & Total amount of ctDNA in bloodstream\\ & at time $t$ \\
$\lambda$ & Shedding rate of malignant cells \\
$\lambda_{\text{bn}}$ & Shedding rate of benign cells \\
$\varepsilon$ & Elimination rate of ctDNA from bloodstream \\
$r = b - d$ & Net growth rate of malignant cells \\
$\Lambda$ & Expected value of $C_t^B$ under null hypothesis \\
$\Delta t$ & Time interval between consecutive observations \\
$S_{ij}(X)$ & $(i,j)$-th second-order Signature coefficient of\\& signal $X$ \\
$LA(X)$ & Lévy area of the path $X$ (anti-symmetric part of\\& $S^{(2)}$) \\
\hline
\end{tabular}
\end{center}
\end{tcolorbox}

\section{A stochastic model for the dynamics of tumorous biomarkers}
In this first section, we summarise past discoveries in the modelling of tumor growth and ctDNA shedding using Birth and Death Processes (BDP). Stochastic models such as BDP are playing an important rôle in capturing the great variability of the signals encountered in observing the evolution of biomarker levels in patients \cite{durrett2015branching}. Based on modelling with BDP's, the dynamics of CtDNA biomarkers in the blood as a function of time has been studied in great depth in \cite{avanzini2020mathematical}. The dynamics of any Birth and Death processe $(X_t)_{t\in \mathbb R_+}$ is governed by the following rules: the process has the Markov property and $P_{i,j}(t)=\mathrm{P}\{X_{t+s}=j \mid X_s=i\}$ describes how $X_t$ changes through time. For small time increments $d t>0$, the function $P_{i, j}(dt)$ is assumed to satisfy the following properties:
\begin{align}
& P_{i, i+1}(\delta t)=b_i \delta t+o(\delta t), \quad i \geq 0 \\
& P_{i, i-1}(\delta t)=d_i \delta t+o(\delta t), \quad i \geq 1, \\
& P_{i, i}(\delta t)=1-\left(b_i+d_i\right) \delta t+o(\delta t), \quad i \geq 1
\end{align}
where $b_i$ is called the birth rate at level $i$ and $d_i$ is called the death rate at level $i$, $i\in \mathbb N$. All the relevant background on Markov processes relevant to our study can be found in \cite{durrett2015branching}. When $b_i$ and $d_i$ do not depend on $i$, the index $i$ is omitted.

\subsection{Biomarker dynamics}
When tumorous cells are present in the body, their death through apoptosis have a certain probability of releasing a specific biomarker, called circulating DNA, or ctDNA, in the bloodstream. In this section, we present the tumor growth model and the resulting ctDNA biomarker shedding process as analysed in \cite{avanzini2020mathematical}. We distinguish between different types of dynamics: the malignant tumor model, the benign tumor model and the biomarker shedding model. 

\textbf{Malignant tumors:} 
             The primary tumor grows stochastically from a single malignant cell at time $t=0$. 
             The tumor size at time $t$, denoted $A_t$ and expressed in number of cells, is modeled by a supercritical branching birth death process with net growth rate 
            \begin{align*}
                r & =b-d>0,
            \end{align*} 
            where $b$ and $d$ denote the birth and death rate per day, respectively. This process has a probability of $\delta=\min\{1,d / b\}$ to go extinct. See e.g. Durrett \cite{durrett2015branching}.
             
Cells in benign tumors and expanded subclones often harbor the same cancer-associated mutations as cancer cells. 
     Hence, ctDNA shed from benign tumors can be difficult to distinguish from ctDNA shed from malignant tumors. 
    
\textbf{Benign tumors:}
        Normal or benign tumor cells also shed cell-free DNA (cfDNA) into the bloodstream. 
         Let $B_t$ denote the size of the benign population of cells at time $t$ that shed the biomarker in the bloodstream, and denote their shedding rate as $\lambda_{\text {bn }}$. We assume that benign lesions roughly replicate at a constant size. Hence, benign cells divide and die at the same rate 
        \begin{align*}
        b_{\mathrm{bn}}=d_{\mathrm{bn}},
    \end{align*}
    for simplicity, we  assume that their population size $B_t$ remains constant over time $\left(B_t=B_0\text{ for all } t\right)$.
    
    Considering again that the biomarker is exclusively shed by cells undergoing apoptosis, the shedding rate of benign cells is 
    \begin{align*}
        \lambda_{\mathrm{bn}}=d_{\mathrm{bn}} \cdot q_{d, \text { bn }}
    \end{align*} per day. 
    
\begin{figure}[!ht]
\begin{center}
\begin{tikzpicture}[scale=1, every node/.style={scale=1}]
    \node[draw=black, fill=red!40, minimum size=1cm, drop shadow] at (0, 0) (cell1) {Tumor Cell};
    
    \draw[->, thick, color=purple] (cell1) -- ++(-3, -1.5) 
        node[midway, above, sloped, font=\small] {Apoptosis};
        
    \node[draw=black, fill=orange!40, minimum size=1cm, drop shadow] at (-2.5, -2.15) (ctDNA1) {ctDNA Shedding};
    \draw[->, thick, color=purple] (ctDNA1) -- ++(0, -1.3);
    
    \draw[fill=blue!20] (-5, -3.5) rectangle (5, -4) 
        node[pos=0.5, above, font=\small] {};
    
    \draw[->, thick, color=purple] (cell1) -- ++(3, 0) 
        node[midway, above, font=\small] {Division};
        
    \node[draw=black, fill=red!40, minimum size=1cm, drop shadow] at (4.5, 0) {Tumor Cell};
    \node[draw=black, fill=red!40, minimum size=1cm, drop shadow] at (7, 0) {Tumor Cell};
    
    \node at (0.5, -3.5) [below, align=center,font=\small\bfseries] {Bloodstream};
    
    \draw[->, thick, color=purple] (0, -4) -- ++(0, -1) 
        node[midway, right, font=\small] {$\epsilon$ shed out};
\end{tikzpicture}
\end{center}
\caption{The shedding process in the case of apoptosis \label{apop}}
\end{figure}
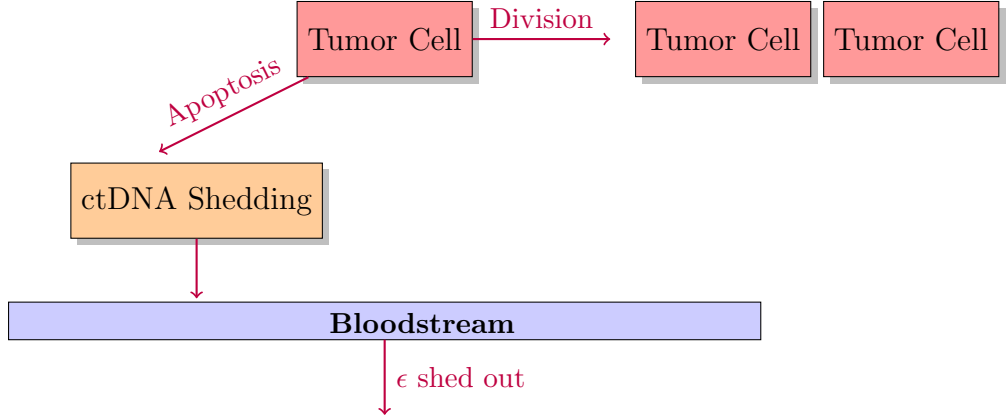

\textbf{Biomarker shedding:}
     Although apoptotic benign cells shed ctDNA with the same probability as malignant cells $\left(q_d=q_{d, \mathrm{bn}}\right)$, we will assume that the shedding rate $\lambda_{\mathrm{bn}}$ of benign cells is lower than the shedding rate $\lambda$ of malignant cells because benign cells typically replicate at a lower rate than cancer cells.
    The circulating biomarker is eliminated from the bloodstream at an elimination rate $\varepsilon$ which can be calculated from the biomarker half-life time $t_{1 / 2}$ as $\varepsilon=\log (2) / t_{1 / 2}$. We denote as $C_t^A$ and $C_t^B$ the amount of biomarker (i.e., number of hGE) circulating in the bloodstream at time $t$ shed by malignant and benign cells,
respectively (we also note $A_t$ and $B_t$ the total amount of cells in, respectively, malignant and benign tumor). The total amount of the biomarker circulating at time $t$ is thus $C_t=C_t^A+C_t^B$. Since malignant and benign cells shed the biomarker independently from each other, the processes $\left(A_t, C_t^A\right)$ and $\left(B_t, C_t^B\right)$ can be studied separately. The stochastic process $\left(A_t, C_t^A\right)$ is a two-type branching process governed by the following transitions
\begin{align}
A \longrightarrow A A \text{ with rate } \quad b, \hspace{1cm} &
A \longrightarrow C \text{ with rate } \quad d \cdot q_d \\
A \longrightarrow \emptyset \text{ with rate } \quad d\left(1-q_d\right), \hspace{1cm} & 
C \longrightarrow \emptyset  \text{ with rate }\quad \varepsilon 
\end{align}
where $q_d$ is defined as the probability of ctDNA release at cell death. The process is initialized at time $t=0$ with a single cancer cell and no circulating biomarkers, that is $\left(A_0, C_0^A\right)=(1,0)$ (see Fig. \ref{apop}).

Since we assume that the benign cell population $B_t$ remains constant over time and biomarker units are eliminated from the bloodstream independently at the same rate, the process $C_t^B$ is a branching pure-death process with constant immigration. We assume that $C_t^B$ is at equilibrium at time $t=0$. 

\textbf{The $C_t^A$ process:}
The biomarker is shed by cancer cells as a Cox process $C^A_t$ (or doubly stochastic Poisson process with intensity $\left(\lambda \cdot A_t\right)_{t \geq 0}$. Furthermore, for large times 
$$\lim _{t \rightarrow \infty} A_t=W e^{r t}\quad 
\text{with }W \stackrel{d}{=} \sum_{i=1}^{A(0)} \chi_i \psi_i$$
where $\chi_i \sim \operatorname{Bern}\left(r /b\right)$ and  $\psi_i \sim \operatorname{Exp}\left(r / b\right)$ are independent. 
 $W=0$ if and only if the process $A_t$ goes extinct. After some generating functional manipulations, we get for large times that
$$C_t^A  \stackrel{\mathcal D}{\approx} \text{Geom}\left(r(\epsilon +r)/(b\lambda e^{rt}+r(\epsilon+r)) \right).$$
    
\textbf{The $C_t^B$ process:} The $C_t^B$ process is a branching pure-death process with immigration, with death rate $\varepsilon$ and immigration rate 
$$
B_0 \cdot \lambda_{\mathrm{bn}}=B_0 \cdot d_{\mathrm{bn}} \cdot q_d,$$ because the population size $B_t$ is constant over time. The probability generating function for such a process is given by 
$$
\mathcal{C}^B(y, t)=\left(1+(y-1) e^{-\varepsilon t}\right)^{C_0^B} e^{\frac{\lambda_{\mathrm{bn}}}{\varepsilon} B_0(y-1)\left(1-e^{-\varepsilon t}\right)} .
$$
When $t$ is large, 
$\lim _{t \rightarrow \infty} \mathcal{C}^B(y, t)=e^{\frac{\lambda_{\mathrm{bn}}}{\varepsilon} B_0(y-1)}.$ 
independently of the initial biomarker level. The right hand side is the probability generating function of a Poisson random variable with mean $B_0 \cdot \lambda_{\mathrm{bn}} / \varepsilon$, and so for large $t$ we have 
\begin{align}C_{t}^B \sim \text { Poisson }\left(\frac{B_0 \cdot \lambda_{\mathrm{bn}}}{\varepsilon}\right).
\label{ctb}
\end{align}
We can assume that $\lambda_{\mathrm{bn}}=\varepsilon$ and $B_0=\mathbb{E}_{t\in\mathbb{R}_+}[C_t^B]$.

\textbf{The $C_t$ process and sampling:}
     By combining the previous results, we find that the asymptotic limit for the distribution of the total ctDNA biomarker amount present in the bloodstream when the primary tumor is made of a large number of cells is the sum of a Cox process and a Death process with immigration, with unknown parameters. 

Based on this model, our plan is to devise a new test for detecting malignant behavior from only a few samples. The main tool we will use is Signature theory, which allows to build a transform that accomodates irregularly sampled observations, as is the case in ctDNA level monitoring. Using Signatures, one circumvents the need to estimate the parameters of the BDP based on too few observations, an impossible statistical challenge.

\section{The Signature transform}
Signature Theory is a set of tools for analysing multidimensional signals, multivariate time series, of the form
\begin{align*}
X(t)=(X^1(t),\ldots,X^d(t)), \end{align*} 
$t\in [0,T]$, which may not be regularly sampled in time and for which the sampling times are not necessarily synchronized across its $d$ dimensions. Signatures are multiple integrals of the product of the derivatives of each component, each with its own variable, and the variables of integration are constrained by a certain given order, which ensures that the speed of certain component is multiplied by the speed of another component at past times only or at future times only. The Signatures come naturally from the expansion of solutions to forced ODE's using the Picard iteration method, but in the 60's, very interesting structures were discovered by Chen and others about Signatures, making them a very interesting tool for mathematicians, data scientists and engineers as well. Several very good references on Signature Theory are \cite{chevyrev2016primer}, \cite{lyons2022signature}, \cite{cass2024lecture}, as well as the first chapter from \cite{fermanian2021learning}. The formal definition of Signatures is given below. 
\begin{defi}[$k^{\text{th}}$-order Signature]
    The signature of order $k$, denoted by \begin{align}
    S^{(k)}_{[0,t]}(X)\in \mathbb R ^{\overbrace{d\times d\times ...\times d}^{k\text{ times}}}=(\mathbb{R}^d)^{\otimes k}
\end{align}  is defined for every word $i_1i_2...i_k$ from $\{1,...,d\}$  by
\begin{align}
    S(X)_{0, t}^{i_1, \ldots, i_k}=\int_{0<t_k<t} \ldots \int_{0<t_1<t_2}  \frac{dX^{i_1}}{dt_1}(t_1) \ldots \frac{dX^{i_k}}{dt_k}(t_k). 
    \label{sigdef}
\end{align}
\end{defi}
When the signal speeds $\frac{dX^{i_j}}{dt_j}(t_j)$, $j=1,\ldots,k$ do not exist, more general definitions allow to circumvent this purely technical difficulty \cite{chevyrev2016primer}. An interesting class of paths to consider for which Signatures can be defined is the space of bounded variation paths from $[0,T] \subset\mathbb{R}$ in $\mathbb{R}^d$, denoted by $\mathcal{BV}(\mathbb{R}^d)$, i.e. all path $X$ such that 
\begin{align}
\|X\|_{\mathcal{BV}} & :=\sup\limits_{D\subset J}\sum\limits_{\begin{array}{rcl}t_i&\in D\\ i&\neq 0\end{array}}\|X_{t_i}-X_{t_{i-1}}\|_2<\infty.
\end{align}
Signatures of order 2 are matrices and signatures of order 3 and larger belong to the type of mathematical objects called tensors \cite{lim2021tensors}, \cite{deisenroth2020mathematics}. Given the definition of all $k^{\text{th}}$-order Signatures, Signatures are now defined as the list of all signatures of all orders, arranged by increasing order, associated with a given multidimensional signal.
\begin{defi}[Signature]
    The \textbf{signature} of a path $X:\mathbb{R}\rightarrow\mathbb{R}^d$ over $I\subset \mathbb{R}$ is an infinite sequence of tensors defined by 
    \begin{align*}
        S_I(X)=\bigotimes\limits_{i=0}^\infty S^{(i)}_I(X)\in\bigoplus\limits_{d=1}^\infty R^{\otimes d} =: T(\mathbb{R}^d).
    \end{align*}
\end{defi}

    Signatures have very useful properties that make them ideal for feature extraction. Signatures would be of poor interest if not coming with very interesting properties allowing the space of Signatures to be endowed with very powerfull structures. 
\begin{itemize}
    \item The first property is \textbf{Invariance to Reparametrisation}, i.e. 
\begin{align} S_I(\tilde X) & = S_I(X)
\end{align}
for all $t \in [0,T]$ with $\tilde{X}(s)=X_{\phi(s)}$ for $\phi$ any surjective, increasing differentiable function $\phi: [0,T]\mapsto [0,T]$. This property is very simple and intuitive as it means that only the trajectory as a curve is preserved but not the speed at wich it is traversed. For instance, the digit "3" is a curve in a 2-dimensional space, but Signatures forget how this digit actually came to be written. For an illustration, see Fig. \ref{reparam}
\begin{figure}
    \centering
    \includegraphics[width=\linewidth]{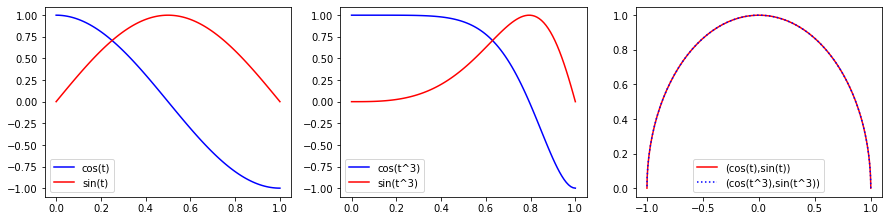}
    \caption{Illustration of reparametrization invariance.}
    \label{reparam}
\end{figure}
\item Another very important property is the 
\textbf{Shuffle Product Identity}. Consider a path $X:[0, T] \mapsto \mathbb{R}^d$ and two multi-indexes $I=\left(i_1, \ldots, i_k\right)$ and $J=\left(j_1, \ldots, j_l\right)$ with $i_1, \ldots, i_k, j_1, \ldots, j_l \in\{1, \ldots, d\}$ with possible multiple repetitions. Notice that $S(X)_{0, T}^I$ (resp.  $S(X)_{0, T}^J$) is a real number that is the corresponding components of the $k^{th}$-order Signature tensor (resp. $l^{th}$-order Signature tensor). The Shuffle Product of two index sets $I$ and $J$, denoted by $I\shuffle J$, can be easily defined after considering the following examples: 
\begin{align*}
\{1,2\} \shuffle \{3\} & =\{\{1,2,3\},\{1,3,2\},\{3,1,2\}\} \\
\{1,2\} \shuffle \{3,4\} & =\{\{1,2,3,4\},\{1,3,2,4\},\{1,3,4,2\},\{3,1,2,4\},\\
&\{3,1,4,2\},\{3,4,1,2\}\} \\
\{1,2\} \shuffle \{2,1\} & =\{\{1,2,2,1\},\{1,2,2,1\},\{1,2,1,2\},\{2,1,2,1\},\\
&\{2,1,1,2\},\{2,1,1,2\}\}.
\end{align*}
The Shuffle Product is more generally defined as all the ways of merging two index sets while still preserving the order of appearance for each initial set in the result set.  More rigorously, any shuffle product can be obtained recursively from the two following basic properties 
\begin{align}
    I \shuffle \emptyset & = I \\
    Ia \shuffle Jb & = \left\{(I\shuffle Jb) a,(Ia\shuffle J)b\right\} 
\end{align}
where $Ia$ is the concatenation of the index set $I$ with the singleton $\{a\}$ and $Jb$ is the concatenation of the index set $J$ with the singleton $\{b\}$. Using these definitions, the Shuffle Product Identity is the following identity
\begin{align}
S(X)_{0, T}^I \cdot S(X)_{0, T}^J & =\sum_{K \in I \shuffle J} S(X)_{0, T}^K.
\end{align}
An illustration of Shuffle product applied to level 2 signature can be found in Fig. \ref{shuff_im}

\begin{figure}
    \centering
    \includegraphics[width=\linewidth]{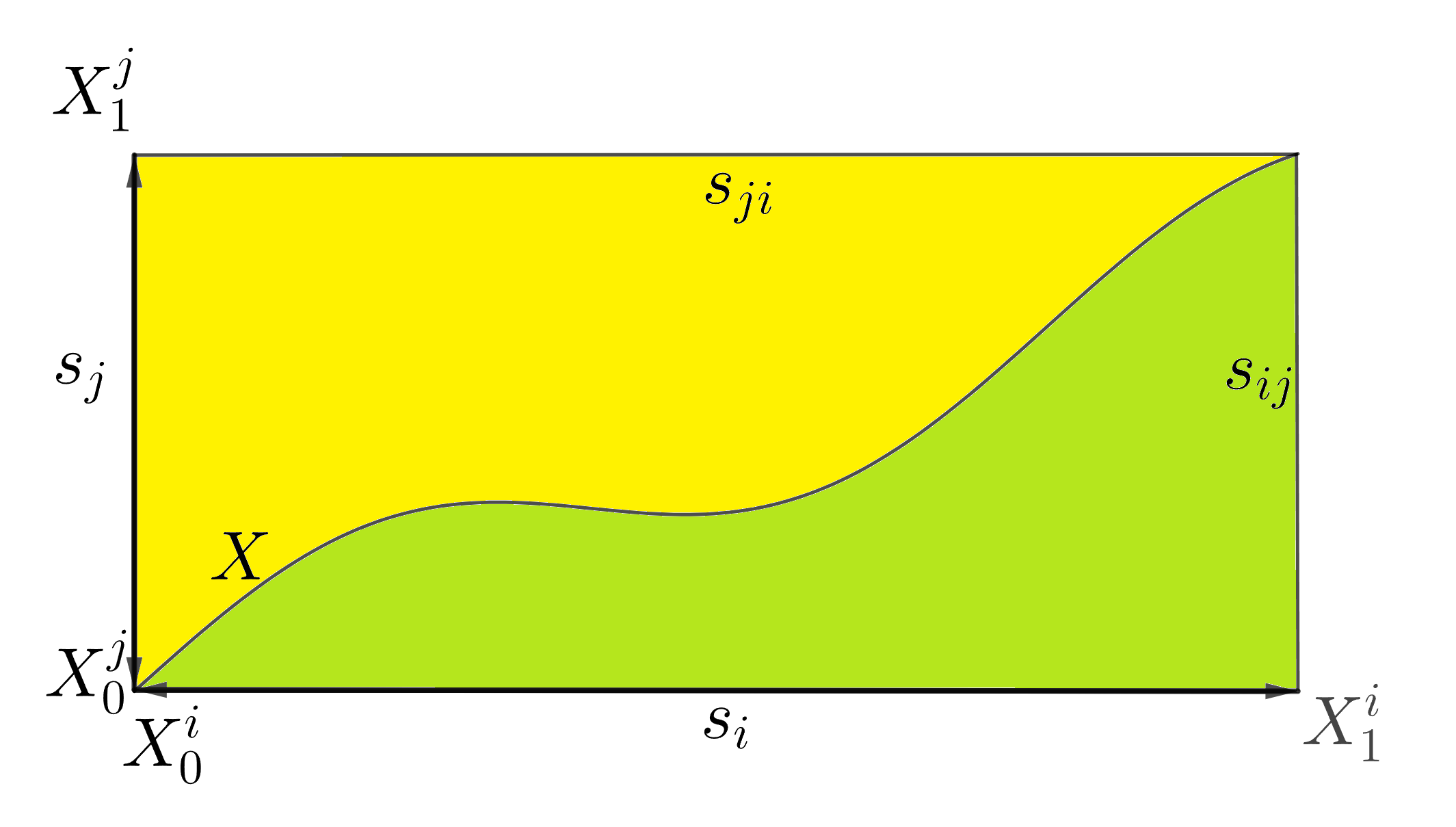}
    \caption{Geometric illustration of shuffle product $S^iS^j=S^{ij}+S^{ji}$.}
    \label{shuff_im}
\end{figure}
\item Finally, \textbf{Chen's Property} completes the presentation of the most elementary properties of Signatures. Let $X:[0, T] \mapsto \mathbb{R}^d$ and $Y:[T, T'] \mapsto \mathbb{R}^d$ be two paths. Define the concatenation of $X$ and $Y$ by 
\begin{align}
    X*Y(t) & = 
    \begin{cases}
        X(t) \quad \text{ for } t\in [0,T] \\
        Y(t) \quad \text{ for } t\in [T,T'].
    \end{cases}
\end{align}
Then, for the concatenation of $X$ and $Y$, we have
\begin{align}
S(X * Y)_{0, T'}=S(X)_{0, T} \otimes S(Y)_{T, T'}.
\end{align}
\end{itemize}
Chen's identity is often used when modelling the signal using piecewise affine. Given a path $X:[0,T]\mapsto \mathbb R^d$ and a set of timestamps $T_\nu=\{t_1,\ldots,t_\nu\}$, one defines $X_{T_\nu}$ as the piecewise linear path obtained by linearly interpolating the observed values of $X$ at times $t_1,\ldots,t_\nu$. Then it is quite easy to compute the Signature of the path using Chen's identify:
\begin{align*}
    S(X_{t_\nu})&=\bigotimes\limits_{i=1}^{\nu-1} S(X)_{t_{i+1},t_{i}}.
\end{align*}

In our work, we will employ the coefficients of first and second degree of the signature, as well as the \textbf{Levy Area}. 
\begin{defi}
    In the case $d=2$, the Levy Area is given by:
\begin{align*}
    A_I(X)=\frac{1}{2}\left( S_I^{12}(X)-S_I^{21}(X)\right).
\end{align*}
\end{defi}
Geometrically, the Levy area computes the area enclosed by a path $X$ on each side of the $[X(0),X(T)]$ segment. The Signatures of order $k$ and the Levy Area will be instrumental for building the detection scheme.

\section{Our statistical testing procedure}
\subsection{Hypothesis testing via signature coefficients}
 
Under the assumption that no malignant tumor is present, we can make the assumption that the expected level of ctDNA is constant. Moreover, assuming that blood samples for a single patient are taken at sufficient distant times, we can assume with a small approximation error that $C^B_t$ has reached its large time distribution (see \ref{ctb}) at $t=0$ and that $C^B_{t_i}$ is independent of $C^B_{t_{i-1}}$. For the sake of simplicity, we will set 
\begin{align*}
    \Lambda & = \mathbb{E}[(C_t^B)_{t\in[0;T]}].
\end{align*}
Consider $X_t = (t,C^B_t)$. According to the usual framework of signature computation, we preprocess the signal. Let denote $\{t_0,t_1,...,t_n\}$ the sampling times of $X$, the signals are centered w.r.t. times:
\begin{align*}
        (\tilde{C^B_t})_{t\in \{t_0,...,t_n\}}\leftarrow (C^B_t)_{t\in \{t_0,...,t_n\}}-\mathbb{E}[(C^B_t)_{t\in \{t_0,...,t_n\}}].
\end{align*}
Another precomputation consists in reducing $[0;T]$ to $[0,1]$, so that $T=1$.

\subsubsection{Construction of the pivotal variable:}
We begin by computing the distribution of the Signature coefficients up to level 2. The first Signature tensor is 
\begin{align*}
    \left(\int\limits_{t=0}^Tdt=T, \hspace{0.5cm} \int\limits_{t=0}^TdC^B_t=C^B_T-C^B_0\right).
\end{align*}
We suppose $C^B_0\sim \text{ Poisson}_c \left(0,\Lambda\right)$ \footnote{Where Poisson$_c$ denotes the \textbf{centered Poisson$(\Lambda)$ distribution}, i.e. the law of $Y_c$ with  $Y_c=Y-\lambda$, with $Y\sim\text { \textbf{Poisson}}\left(\lambda\right)$}.
Based on this, we deduce that
\begin{align}
S^{2}(X) =C^B_T-C_0^B\sim \text{\textbf{Skellam}}(\Lambda,\Lambda).
\label{s2}
\end{align}

giving us the distribution of our first pivotal variable\footnote{Recall that we consider $\Lambda = \mathbb{E}[(C_t^B)_{t\in [0;T]}]$}.

Let us now address the question of computing the  2nd order signature coefficients for the signal $X_t$. 

The distribution of the integral of $t$ with respect to the increments of a Poisson process $C_t^B$, denoted $dC_t^B$, sampled in $\sigma_1 = \{0=t_0,...,t_n=1\}$, is given by 
\begin{align}
S^{12}_{[0,T]}(X)= T(C^B_T-C_0^B)-S^{21}_{[0;T]}(X) \label{shuffle},
\end{align}
using the shuffle product $S^1S^2=S^{12}+S^{21}$ (see \cite{chevyrev2016primer} for the definition of the Shuffle product and its relationship with Signatures). In order to compute $ S^{12}(X)$, we use a linear interpolation over $\sigma_1$. To simplify calculation, we assume that $t_{i+1}-t_i=\Delta t$ for all $i\in\{0,...,n\}$. First, as we consider a linear interpolation over $\sigma_1$, we have 
\begin{align*}
    C_t^B=\sum\limits_{i=0}^{n-1}\left(\frac{C_{t_{i+1}}-C_{t_i}}{\Delta t}t+cst\right)\mathbb{1}_{[t_i;t_{i+1}]}(t),
\end{align*}
and then,
\begin{align}
    dC_t^B=\frac{C_{t_{i+1}}^B-C_{t_i}^B}{\Delta t}dt.
    \label{dctb}
\end{align}
It follows
    \begin{align*}
        S^{12}_{[0;T]}(X)=\int\limits_0^TtdC_t^B&=\sum\limits_{i=0}^{n-1}\int\limits_{t_i}^{t_{i+1}}tdC_t^B\\
        \text{(using \ref{dctb})}&= \sum\limits_{i=0}^{n-1}\int\limits_{t_i}^{t_{i+1}}\frac{C_{t_{i+1}}^B-C_{t_i}^B}{\Delta t}tdt= \frac{1}{2}\sum\limits_{i=0}^{n-1}\frac{C^B_{t_{i+1}}-C^B_{t_i}}{\Delta t}(t_{i+1}^2-t_i^2)\\
        &= \frac{1}{2}\ \sum\limits_{i=0}^{n-1}\left(C^B_{t_{i+1}}-C^B_{t_i}\right)(t_{i+1}+t_i)= \frac{1}{2}\sum\limits_{i=0}^{n-1}\left(C^B_{t_{i+1}}-C^B_{t_i}\right)(2t_i+\Delta t)\\
        &= \sum\limits_{i=0}^{n-1}t_i\left( C^B_{t_{i+1}}-C^B_{t_i}\right)+\frac{\Delta t}{2}\sum\limits_{i=0}^{n-1}\left(C^B_{t_{i+1}}-C^B_{t_i}\right).
    \end{align*}
Reorganizing the sums and using that $t_i+\Delta t = t_{i+1}$, we get
\begin{align*}
    S^{12}_{[0;T]}(X)&=\sum\limits_{i=0}^{n-1}t_{i+1}C_{t_{i+1}}^B-\sum\limits_{i=0}^{n-1}t_iC^B_{t_i}-\Delta t \sum\limits_{i=0}^{n-1}C_{t_{i+1}}^B+\frac{\Delta t}{2}\sum\limits_{i=0}^{n-1}\left(C^B_{t_{i+1}}-C^B_{t_i}\right).
\end{align*}
Since
\begin{align*}
\sum\limits_{i=0}^{n-1}t_{i+1}C_{t_{i+1}}^B=\sum\limits_{i=1}^{n}t_iC_{t_i}^B 
\end{align*}
and
\begin{align*}
    -\Delta t \sum\limits_{i=0}^{n-1}C_{t_{i+1}}^B+\frac{\Delta t}{2}\sum\limits_{i=0}^{n-1}\left(C^B_{t_{i+1}}-C^B_{t_i}\right)=-\frac{\Delta t}{2}\left(\sum\limits_{i=0}^{n-1}C_{t_{i+1}}^B+\sum\limits_{i=0}^{n-1}C_{t_i}^B\right),
\end{align*}
it follows that
\begin{align*}
    S^{12}_{[0;T]}(X)&=TC_T^B-\underbrace{t_0C_0^B}_{=0}-\frac{\Delta t}{2}\sum\limits_{i=0}^{n-1}\left( C_{t_{i+1}}^B+C_{t_i}^B\right).
\end{align*}
We conclude that $S^{12}_{[0;T]}(X)$
\begin{align*}
    S^{12}_{[0;T]}(X)&=TC_T-\frac{\Delta t}{2}\left(\sum\limits_{i=0}^{n-1}C_{t_{i+1}}^B+\sum\limits_{i=0}^{n-1}C_{t_i}^B\right)\\
    &= \left(T-\frac{\Delta t}{2}\right)C^B_T-\frac{\Delta t}{2}C^B_{t_0}-\Delta t\sum\limits_{i=1}^{n-1}C^B_{t_i}.
\end{align*}
We can now recover the distribution of $S_{(0,T]}^{12}(X)$, under the assumption that $T\in\mathbb{N}$ and $\Delta t\in 2\mathbb{N}$:
\begin{align*}
    S^{12}_{[0;T]}(X)=\underbrace{\left(T-\frac{\Delta t}{2}\right)C_T^B}_{\sim \mathcal{P}\left((T-\frac{\Delta t}{2})\Lambda\right)}-\underbrace{\left(\underbrace{\Delta t\sum\limits_{i=1}^{n-1}C_{t_i}^B}_{\sim \mathcal{P}\left((n-1)\Delta t\Lambda\right)=\mathcal{P}((T-\Delta t)\Lambda)}+\underbrace{\frac{\Delta t}{2}C_{t_0}^B}_{\sim \mathcal{P}\left(\frac{\Delta t}{2}\Lambda\right)}\right)}_{\sim \mathcal{P}\left((T-\frac{\Delta t}{2})\Lambda\right)}.
\end{align*}
As we saw, difference between two Poisson distribution is a Skellam distribution, so:
\begin{align}
    S^{12}_{[0;T]}(X)\sim \textrm{\textbf{Skellam}}(T\Lambda,T\Lambda).
    \label{s12}
\end{align}
Using \ref{shuffle}, we recover $S^{21}_{[0;T]}(X)$ from $S^{12}_{[0;T]}(X)$:
\begin{align*}
    S^{21}_{[0;T]}&=TC_T^B-TC_0^B-S^{12}_{[0;T]}(X)\\
    &= TC_T^B-TC_0^B-\left(T-\frac{\Delta t}{2}\right)C_T^B+\frac{\Delta t}{2}C_0^B+\Delta t\sum\limits_{i=1}^{n-1}C_{t_i}^B\\
    &= \underbrace{\underbrace{\frac{\Delta t}{2}C_T^B}_{\sim \mathcal{P}\left(\frac{\Delta t}{2}\Lambda\right)}+
    \underbrace{\Delta t\sum\limits_{i=1}^{n-1}C_{t_i}^B}_{\sim \mathcal{P}\left((T-\Delta t)\Lambda\right)}}_{\sim \mathcal{P}\left((T-\frac{\Delta t}{2})\Lambda\right)}-
    \underbrace{\left(T-\frac{\Delta t}{2}\right)C_0^B}_{\sim \mathcal{P}\left(\left(T-\frac{\Delta t}{2}\right)\Lambda\right)}.
\end{align*}
In conclusion for $S^{21}_{[0;T]}(X)$
\begin{align}
    S^{21}_{[0;T]}(X)\sim \textrm{\textbf{Skellam}}\left(\left(T-\frac{\Delta t}{2}\right)\Lambda,\left(T-\frac{\Delta t}{2}\right)\Lambda\right).
    \label{s21}
\end{align}
At last, we can compute $\mathcal{LA}_{[0;T]}(X)$:
\begin{align*}
    2\mathcal{LA}_{[0;T]}(X)&=S^{12}_{[0;T]}(X)-S^{21}_{[0;T]}(X)\\
    &=2S^{12}_{[0;T]}(X)-TC_T^B+TC_0^B\\
    &= 2\left(\left(T-\frac{\Delta t}{2}\right)C^B_T-\frac{\Delta t}{2}C^B_{t_0}-\Delta t\sum\limits_{i=1}^{n-1}C^B_{t_i} \right)-TC_T^B+TC_0^B\\
    &= \left(2T-\Delta t\right)C^B_T-\Delta tC^B_{t_0}-2\Delta t\sum\limits_{i=1}^{n-1}C^B_{t_i}-TC_T^B+TC_0^B\\
    &=
    \underbrace{\underbrace{(T-\Delta t)C_0^B}_{\sim \mathcal{P}((T-\Delta t)\Lambda)}+
    \underbrace{(T-\Delta t)C_T^B}_{\sim \mathcal{P}((T-\Delta t)\Lambda)}}_{\sim \mathcal{P}(2(T-\Delta t)\Lambda)}-\underbrace{2\Delta t\sum\limits_{i=1}^{n-1}C_{t_i}^B}_{\sim\mathcal{P}(2(T-\Delta t)\Lambda)}.
\end{align*}
Finally, dividing by 2
\begin{align}
    \mathcal{LA}_{[0;T]}(X)\sim \textrm{\textbf{Skellam}}\left((T-\Delta t)\Lambda,(T-\Delta t)\Lambda\right).
    \label{la}
\end{align}
We did all computations considering $T\in\mathbb{N}$ and $\Delta t\in 2\mathbb{N}$. These hypothesis fits a "real" context where timestamps are days, timesteps are long enough to be considered even, and allows to sum Poisson distributions.
To retrieve distributions corresponding to signature framework, we just consider $T=1$ and $\Delta t=\frac{1}{n}$.  

\subsection{Empirical distribution of signature coefficients for benign trajectories.}

\begin{figure}[!ht]
        \centering
            \includegraphics[width=0.8\linewidth]{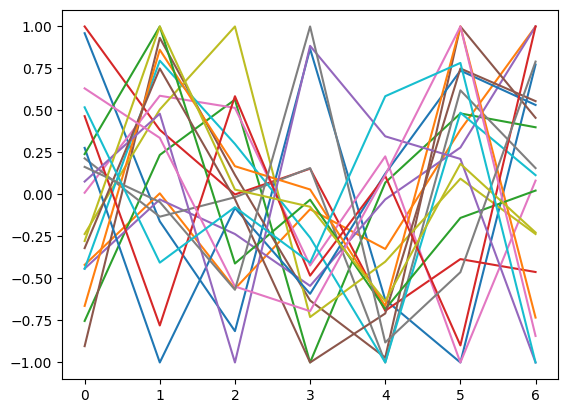}
            \caption{Benign trajectories $C_t^B$ as a function of time.}
            \label{traj1}
\end{figure}

\begin{figure}[!ht]
    \centering
    \includegraphics[width=0.8\linewidth]{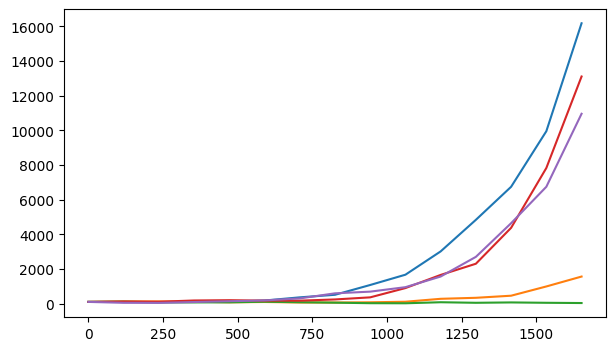}
    \caption{Benign then malignant trajectories $C_t^{BM}$ as a function of time.}
    \label{traj2}
\end{figure}

\begin{figure}[!ht]
    \centering
    \includegraphics[width=0.8\linewidth]{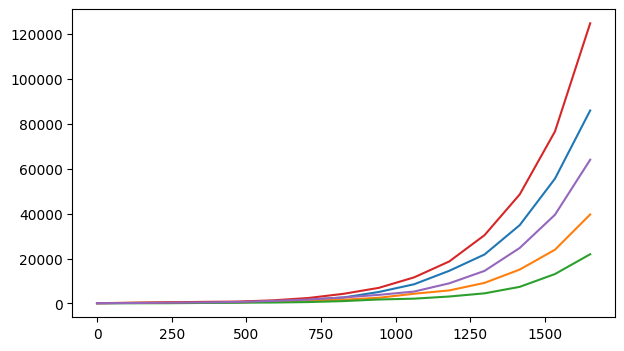}
    \caption{Pure malignant trajectories as a function of time.}
    \label{traj3}
\end{figure}

We sampled 4000 benign ($\lambda\in\Lambda=\{0.1,0.12,0.14,0.17\}$, denoted by $C_t^B$) and 36000 benign then malignant (with a random transition time leading $\lambda\in\Lambda$ and $r=b-d=0$ to become $r\in \mathcal{R}=\{0.001,0.002,0.004,0.007,0.01,0.015\}$, corresponding to an aggressive tumorous dynamic trajectory, denoted by $C_t^{BM}$. For the sake of completeness, we also sampled pure malignant trajectories, noted $C_t^M$, see Fig. \ref{traj1}, \ref{traj2}, \ref{traj3}.

We used the BirDePy python package \cite{hautphenne2021birth} to sample 7 observations times from a continuous trajectory. These times are distributed every 300 days, as coefficients for birth and death models are designed for a day by day evolution. We interpret this experiment as exploring the distribution of simulated dynamics with 1 blood samples every year in a 7 years time range.

We now compare the theoretical approximate distributions established in the former paragraph with the empirical distributions obtained using the dataset. The empirical and theoretical distributions are displayed in Figure \ref{emps2}, \ref{emps12}, \ref{emps21}, \ref{empla} below. These empirical results confirm the adequation between the asymptotical distribution and the observed distribution of the Signature coefficients and the Levy Area. 

\begin{figure}[!ht]
    \centering
    \includegraphics[width=0.8\linewidth]{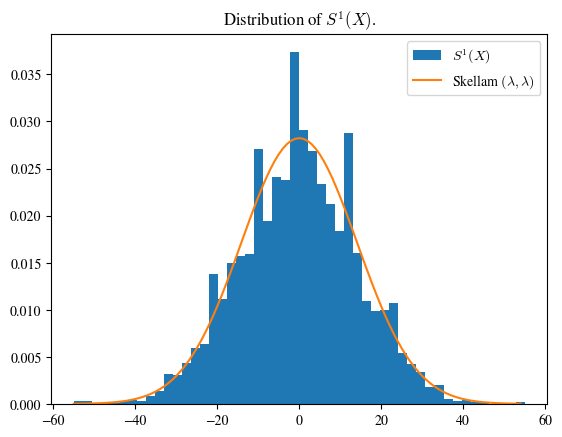}
    \vspace*{1ex}
    \caption{Empirical distribution of $S^1(X)$ against its theoretical distribution from Equation \ref{s2}}
    \label{emps2}
\end{figure}

\begin{figure}[!ht]
\centering
\includegraphics[width=0.8\linewidth]{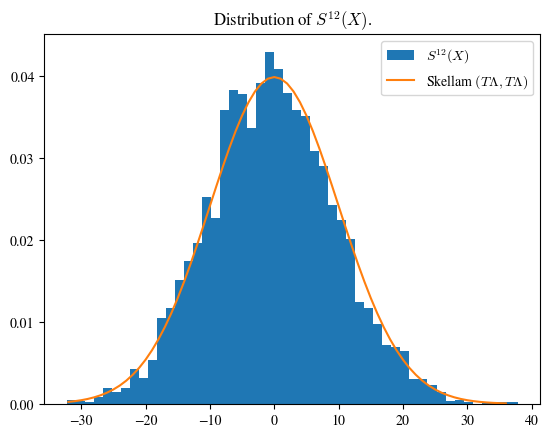}
\vspace*{1ex}
\caption{Empirical distribution of $S^{12}(X)$ against its theoretical distribution from Equation \ref{s12}}
\label{emps12}
\end{figure}

\begin{figure}[!ht]
\centering
\includegraphics[width=0.8\linewidth]{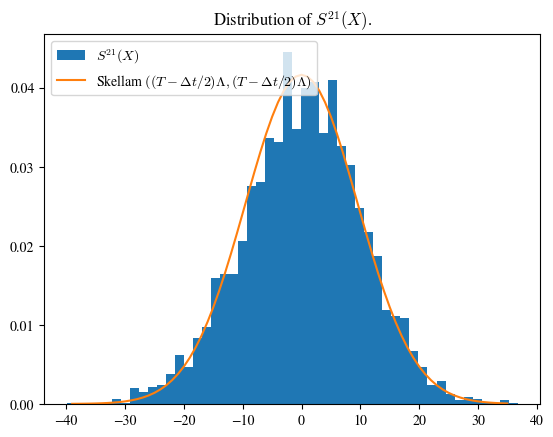}
\vspace*{1ex}
\caption{Empirical distribution of $S^{21}(X)$ against its theoretical distribution from Equation \ref{s21}}
\label{emps21}
\end{figure}

\begin{figure}[!ht]
\centering
\includegraphics[width=0.8\linewidth]{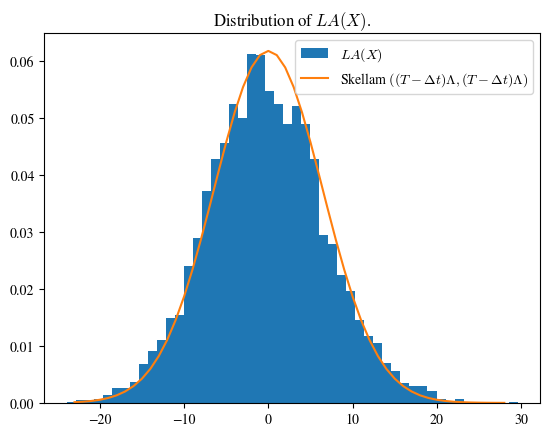}
\vspace*{1ex}
\caption{Empirical distribution of $LA(X)$ against its theoretical distribution from Equation \ref{la}}
\label{empla}
\end{figure}

\subsection{Our testing scheme}
Based on the explicit form of the distribution of our pivotal quantities under the Null Hypothesis, i.e. Benign state of ctDNA dynamics we obtain the testing procedure in Algorithm \ref{algo}. We used a False discovery rate control with a Benjamini-Hochberg correction, as our method is a multiple-testing procedure.

\begin{algorithm}[!ht]
\caption{Testing for $S^w$ with $w\in\{2,12,21,12-21\}$}\label{algo}
\begin{algorithmic}
\Require $X$ and $w\in\{2,12,21,12-21\}$ (with the notation $12-21$ refering to the Levy Area).
\Ensure $(H_0):$ the tumor is benign.
\State Compute $\mathcal C=\frac{1}{n}\sum\limits_{i=0}^nC_{t_i}$ with $n$ the number of observations of $X$.
\State Compute $S^w(\tilde{X})$ with $\tilde{X}=(t,C_t-\mathcal{C})$.
\State Compute the $p$-value corresponding to the observed value of $S^w$ with $\Lambda=\mathcal C$.
\State \hspace{-.3cm}\textbf{Output:} The selected tests for which $H_0$ is not corroborated by the observed data, using the mutliple testing correction proposed by, i.e. Benjamini-Hochberg \cite{benjamini1995controlling}. 
\end{algorithmic}
\end{algorithm}

\subsection{Results}
Our results based on simulated data are collected in the following Table \ref{tab}. 
\begin{table}[t]
\centering
\caption{Scores for each coefficient tests.}

\label{tab1}
\begin{tabular}{|l|l|l|l|l|}
\hline
Test &  Accuracy & Precision & Recall & F1-score\\
\hline
$T_{2}$ &  0.946 & 0.999 & 0.999 & 0.999\\
\hline

$T_{12}$ & 0.945 & 0.999 & 1 & 0.999\\
\hline

$T_{21}$& 0.943 & 0.999 & 0.951 & 0.974\\
\hline

$T_{LA}$ & 0.945 & 0.999 & 0.999 & 0.999\\
\hline

Corrected multiple test & 0.958 & 0.999 & 0.999 & 0.999\\
\hline
\end{tabular}
\label{tab}
\end{table}
These results show that all our tests enjoy a very high detection capability on simulated datasets, a very promising result (accuracy$>0.94$ and F1-score of 0.999). Note that performing accurate detection based on 7 samples only per patient using individual Birth and Death models, with each patient its own coefficients, is statistically extremely difficult to achieve in theory and in practice. This makes our method, to the best of our kniowledge, the first of its kind for early detection on time ordered blood samples under reasonable assumptions about data collection and sample size.

\section{Conclusion and future work}

In this paper, we proposed a new testing procedure based on the Signature Transform for the analysis of scarse blood samples data, which circumvents the statistically impossible task of estimating the coefficients of a Birth and Death Model before testing possible malignancy of a patient's ctDNA level dynamics. The performance of our new testing procedure for early detection of malignant trajectories when applied to a set of simulated trajectories with 7 samples over 2000 days, confirmed the hypothesis that Signatures can extract very significant information about the ctDNA signals without supervised feature learning. 

\vspace{.3cm}

The main future prospects are :

\begin{itemize}
    \item \textit{Confront our method to real data:} Since publication of real medical data is heavily constrained by hard to obtain agreements, we did not present the performance of our method on real data sets in the present paper, but we expect to be able to do so in a near future.
    \item \textit{Compare with competitors:} This paper demonstrates the capabilities of our method but more experiments will be soon available with comparisons with other methods. Notice that we did not find any alternative feature extraction approach capable of approaching the goals of this study. 
    \item\textit{Theoretical studies:} As often the case in medicine, minimizing False Negatives is crucial. We will therefore need to investigate the statistical power of our procedure from a theoretical viewpoint. 
     \item\textit{Lowering to no greater than 5  samples}: 7 samples in a 2000 days period may be considered realistic. However, it could be interesting to lower the number of samples even further, possibly with the same value of $\Delta t$ but a lowering value for $T$.
     \item\textit{Computing the distribution using a random model for $\Delta t$:} Accounting for the randomness is important for better modelling the sampling process. We will devise a bespoke stochastic model in future work.
     \item\textit{Studying the distribution under $(H_0)$ for other coefficients:} We reduced the problem on $S^{(2)}(X)$ in order to simplify theoretical computation. Next step is to generalize these results to every coefficients in $S(X)$.
     \item\textit{Using conformal $p$-values:} can help addressing more general testing problems and should be put to work for our model of tumour growth in order to control the False Detection Rate.
\end{itemize}

\paragraph{\textbf{Acknowledgements.}}
This work was funded by the grant 900R17NEU project "Méthodes neuromorphiques pour la détection précoce de pathologies cancéreuses" from La Région Rhône-Alpes, France. 

\bibliographystyle{elsarticle-num-names} 
\bibliography{biblio.bib}



\end{document}